# Classical analogy of a cat state using vortex light

Shi-Long Liu[1,2], Qiang Zhou[1,3,4], Shi-Kai Liu[1,2], Yan Li[1,2], Yin-Hai Li[1,2], Zhi-Yuan Zhou[1,2], Guang-Can Guo[1,2] & Bao-Sen Shi[1,2]

Cat states are systems in a superposition of macroscopically distinguishable states; this superposition can be of either classically or quantum distinct states, regardless of the number of particles or modes involved. Here, we constructed an experimental model that simulates an optical cat state by engineering the classical orbital angular momentum of light, referred to here as an analogous cat state (a-CS). In our scheme, the behaviors of the a-CS in position space show many similarities to the quantum version of the optical cat state in phase space, for example, movement, rotation, and interference. Experimentally, the a-CS, which has two spatially localized Gaussian intensity lobes, can be evolved from "kitten" to "cat" by engineering the acquired phase hologram. Additionally, we simulated the "decoherence" of the a-CS influenced by atmospheric turbulence. The a-CS provides a reliable tool for visualizing and studying the behaviors of quantum cat states in phase space.

[1] Key Laboratory of Quantum Information, University of Science and Technology of China, Hefei, Anhui 230026, China. [2] Synergetic Innovation Center of Quantum Information & Quantum Physics, University of Science and Technology of China, Hefei, Anhui 230026, China. [3] Institute of Fundamental and Frontier Science, University of Electronic Science and Technology of China, Chengdu 610054, China. [4] School of Optoelectronic Science and Engineering, University of Electronic Science and Technology of China, Chengdu 610054, China. Correspondence and requests for materials should be addressed to Z.-Y.Z. (email: zyzhouphy@ustc.edu.cn) or to B.-S.S. (email: drshi@ustc.edu.cn)

Schrödinger's cat, a famous Gedankenexperiment, devised by Erwin Schrödinger in 1935[1,2], presents an unfortunate cat being both dead and alive inside a sealed box with a single radioactive atom. In analogy with this famous Gedankenexperiment, the modern notion of the Schrödinger cat state is a superposition state of two macroscopically distinguishable states (e.g., the "alive" and "dead" cat states) in a certain degree of freedom (DOF)[3]. Such states lie at the heart of quantum mechanics, not only as features in the theoretical interpretation of the quantum world, but also in quantum information processing, such as quantum computation and metrology[4,5]. In quantum optics, one particularly interesting quantum cat state (q-CS) is realized via a superposition of two coherent states $|\pm\alpha\rangle$ with opposite phase. The q-CS has the form $|CS_\pm\rangle = N\left(|\alpha\rangle_{\text{Alive}} \pm e^{i\theta}|-\alpha\rangle_{\text{Dead}}\right)$, where $N$ is a normalization factor. Over the last few decades, the creation and enlargement of these types of cat states have been studied widely in the photon-number representation $|n\rangle$[6–12]; the generation of photon-subtracted squeezed-vacuum states is one such example[7,9,10]. Unfortunately, most of the preparation mechanisms still produce "small cat" states, also sometimes referred to as "kittens". Experimentally, the manipulation of a "large cat" is very difficult; it is therefore valuable to explore their generation in other representations. Significant advances have been made in several systems[13,14], e.g., excitations of different Rydberg states to fit a cat state[15,16] and formations of cat states by two Bose-Einstein condensates[17–19]. Nevertheless, creating a "cat" state is not an easy task, and the behaviors are so abstract that one has to reconstruct it in a specific space, for example in phase space[7,10,12], energy space[20], or another specific space[14,15,21].

Engineering a cat state in various physical platforms has always been an exciting endeavor. Inspired by the optical cat state $|CS_\pm\rangle$ defined in the photon number (Fock) basis $|n\rangle$, we try to reconstruct a cat-like state in a Hilbert space formed by a spatial mode, i.e., the orbital angular momentum (OAM) of light[22–25]. The pioneering work in 1993 of Allen et al. explored the connection between the high-order modes of the beam in the form of Glauber coherent states[24,26], but surprisingly remained unexplored further. Theoretically, we find that the intensity distribution $I(x, y)$ of the a-CS is similar to the Wigner function $W(x, p)$ of the q-CS, which indicates one can simulate various abstract behaviors of the q-CS in phase space. Notably, the a-CS will be naturally divided into two spatially distinct Gaussian intensity lobes (GILs) in position space via a neat OAM superposition; further, the distance between two GILs increases linearly along with the size of an analogous cat state (a-CS). These behaviors are closely analogous to Schrödinger's macroscopic superposition: the "alive cat" living in a place and the superposition state "dead cat" living in another place.

As an analogy, one can simulate not only the abstract behavior in phase space, but also most characteristics of q-CS, e.g., classifying an a-CS as an even or odd cat state via a photon or mode number occupation. Additionally, one can study interference in a Bloch sphere, and define a seeming "macroscopicity" criterion based on two GILs[27,28]. Nevertheless, one should note that the OAM-based cat states have exactly zero quantum macroscopicity as the state formed by fully classical electronic modes[29]. We only prepare it with classical optical waves to act as an analogy, and the same experiment could be performed with other classical waves in principle, for example, sound or water wave. Studies of the a-CS in the classical world can help one to understand and visualize the behaviors of a "true" cat state in the quantum world. Furthermore, having a source of such a-CSs would represent a new platform on which new cat state-based protocols could be prepared in the future.

Indeed, many significant advances have been made in discovering suitable analogies of quantum phenomena[30–33] using classical waves; typically, a "classical entanglement" is created using spin and OAM degree of freedom (DOF) of light[31,33,34]. Such vortex beams can furnish a good "laboratory" in which to study quantum physics phenomena[32].

In this article, we first propose a way to simulate the q-CS using spatial structured light. Experimentally, using classical OAM light, we successfully generate an a-CS that evolves from a "kitten" $\left(|\alpha_L| = \sqrt{0.5}\right)$ to a "cat" $\left(|\alpha_L| = \sqrt{5.0}\right)$ with high quality. Specifically, we simulate the dynamics of the q-CS completely in phase space, e.g., movement, rotation, and interference; test its coherent property, and evaluate it using standard state tomography. Furthermore, we simulate its "decoherence" by quantifying the effects of atmospheric turbulence on a-CS. For this type of cat state, one can also study its behaviors in a discrete and continuous phase space with a discrete cylinder $S_1 \times \mathbb{Z}$[35,36]. However, because reconstructing such a state in a cylinder is quite complicated in experiments, we do not provide a detailed discussion in this paper. The ability to generate such states not only provides a reliable tool for visualizing the abstract behaviors of q-CSs in phase space, but it may also allow us to explore the physics associated with cat states.

## Results

**Designing an a-CS in the classical world**. An a-CS $|Cat\rangle = N\left(|\alpha\rangle_{\text{Alive}} + e^{i\theta}|-\alpha\rangle_{\text{Dead}}\right)$ encoded in OAM space is a coherent superposition of two distinct opposite-amplitude states, namely $|\alpha_L\rangle$ and $|-\alpha_L\rangle$. One of them, i.e., $|\alpha_L\rangle$, is a weighted superposition of each OAM eigenstate $|L\rangle$, similar to the usual optical coherent state formalism[26,37],

$$|\alpha_L\rangle = e^{-|\alpha_L|^2/2} \sum_{L=0}^{\infty} \frac{\alpha_L^L}{\sqrt{L!}} |L\rangle. \quad (1)$$

Because of the similarity, we refer to it as an analogous coherent state (a-CST). The OAM eigenstate we used is Laguerre-Gaussian (LG) mode: $E_L = A(r, z)\exp(iL\phi)$, where the azimuthal dependence $\exp(iL\phi)$ gives rise to the spiral wavefronts (see Fig. 1a); the fundamental electric field $E_L$ is written as $|L\rangle$ for similarity. In Eq. (1), $\alpha_L (= |\alpha_L|e^{i\varphi})$, a complex number, determines the size and position of the a-CST in position space (see Supplementary Figs. 1, 2, and 3). In particular, with $\alpha_L = 0$, the a-CST $|\alpha_L\rangle = |0\rangle$ signifies a Gaussian beam, i.e., an analogous vacuum state. In the cylindrical coordinate representation, the intensity distribution of the a-CS takes the following form (see Supplementary Note 1),

$$|\langle \text{Cylin}(r, \phi, z)|\text{Cat}\rangle|^2 = G(\alpha_L) + G(-\alpha_L) + 2\cos(\theta - 2\alpha_L \sin(\Phi)) \cdot G(0) \quad (2)$$

where $G(0)$ is the Gaussian beam; $G(\alpha)$ and $G(-\alpha)$ represent two degenerate GILs moving to the right and left, respectively, as the referenced beam $G(0)$. The distance between the centers of the GILs is proportional to the size of $|\alpha_L|$. Furthermore, the third term of Eq. (2) represents the interference between the "alive cat" and "dead cat" states, which is worth discussing for a q-CS[38]. Other parameters are explained in detail in the methods section. The intensity distribution in Eq. (2) for an a-CS does not include topologic charge $L$, but it does carry the spiral spectral information of the OAM modes in a similar manner to the Wigner function of the q-CS in phase space[38–40]:

$$W_{\text{cat}} = W_0(x - x_0, p) + W_0(x + x_0, p) + 2\cos(4x_0 p + \varphi) \cdot W_0(x, p) \quad (3)$$

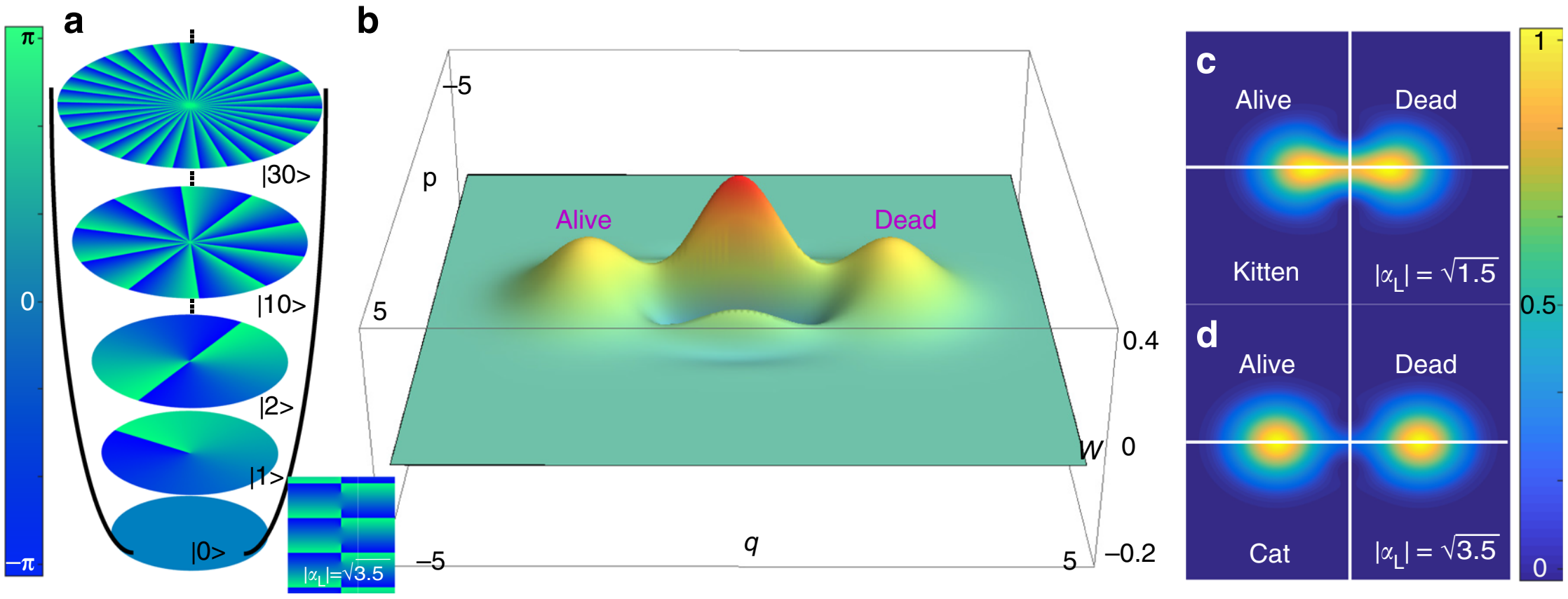

**Fig. 1** Schematics and intensity distributions of an analogous cat state (a-CS). **a** Spiral wavefronts of orbital angular momentum (OAM) of light that forms a Hilbert space. **b** the Wigner function involves two states of a cat state in its "dead" and "alive", where the size of the cat state is $\sqrt{3.5}$. **c**, **d** Intensities of a "kitten" $\left(|\alpha_L| = \sqrt{1.5}\right)$ and "cat" $\left(|\alpha_L| = \sqrt{3.5}\right)$ state, respectively, where $|\alpha_L|$ represents the size of a-CS. An acquired phase hologram for an a-CS is inset in the bottom left of panel **b**. The wavelength and beam waist in all simulations was 780 nm and 200 μm, respectively

Equation 3 represents two Gaussian wave-packets and their interference in phase space (see Fig. 1b). For a q-CS, the size and relative phase factor can completely determine its behaviors in phase space, i.e., movement, rotation, and interference. Correspondingly, the a-CS, $|\mathrm{Cat}\rangle = N\left(|\alpha\rangle_{\mathrm{Alive}} + e^{i\theta}|-\alpha\rangle_{\mathrm{Dead}}\right)$, has similar behaviors in position space (see Fig. 1c, d), i.e., $\alpha_L (= |\alpha_L| e^{i\varphi})$ gives rise to movement ($|\alpha_L|$) and rotation ($e^{i\varphi}$); the relative phase factor $e^{i\theta}$ depicts the interference between two GILs (see a dynamic process in Supplementary Movie 1 and 2). Interestingly, the interference disappears as the a-CS grows, as can be seen in Fig. 1c, d. Additionally, one can define a macroscopicity criterion for the q-CS in phase space via the separation of two Gaussian packets[27,28], i.e., $M = 2|\alpha_L|$ for the cat state in ref. [28]. Similarly, for an a-CS with two GILs in position space, a "macroscopicity" criterion can also be defined as $M = \sqrt{2}|\alpha_L|/2$(see Supplementary Note 2).

In Fig. 1, panels c and d show the intensity distributions of two a-CSs with $|\alpha_L| = \sqrt{1.5}$ (kitten) and $|\alpha_L| = \sqrt{3.5}$ (cat), respectively. Usually, the q-CS can be classified as an even or odd cat based on the occupation number of the photons. Here, the a-CSs can also be sorted based on their occupation on OAM modes. For example, an even a-CS has only even OAM modes $\left(|\mathrm{Cat}\rangle_{\mathrm{Even}} = c_0|0\rangle + c_1|2\rangle + c_2|4\rangle + \ldots\right)$; likewise, the odd a-CS has only odd OAM modes.

The experimental setup for generating an a-CS is shown in Fig. 2a. A 780 nm Ti: sapphire laser was coupled into a single mode fiber as the input. The system has two spatial light modulators (SLM), SLM_G and SLM_M, each with an active area of 12.5 mm × 7.1 mm. SLM_G, placed after the polarization beam splitter (PBS), was used to prepare an a-CS by loading on an acquired blazing-gray phase (the detailed phase hologram is shown in the methods section). SLM_M was located on the focal plane of a 4*f* system consisting of two lenses ($f = 350$ mm); it was used to perform projection measurements for the OAM-based state with the help of a single mode fiber. The overall system has excellent mode orthogonality (see Supplementary Fig. 4). One can observe the behavior of the a-CS directly in position space via a charge coupled device (CCD) camera[41] (see Fig. 2b, c). Our primary and most exciting results from using this set-up are presented in the following, including the spatial mode distributions (spatial spectra), coherence, and state tomography. Furthermore, we theoretically simulated the "decoherence" via the atmospheric turbulence.

**The characteristics of the a-CS**. The normalized spatial spectra for an a-CS with $\alpha_L = \sqrt{0.5}$, $\sqrt{1.5}$, and $\sqrt{3.5}$ are shown in Fig. 3, where the orange and green bars represent the experimental and theoretical normalized results, respectively. The corresponding intensity distributions acquired by the CCD camera are inset in Fig. 3. Obviously, as the a-CS grows, the separation between the two centers of the GILs increases; by measuring the separation distribution, the normalized relative separations from "kitten" to "cat" were found to be 1:2.68:3.30 (1:2.26:3.16 in theory). The overlap between the two GILs will approach zero when the size of $|\alpha_L|$ is equal to $\sqrt{3.5}$, which indicates that the value of $\sqrt{3.5}$ can be used as an effective boundary between analogous "kitten" $\left(|\alpha_L| < \sqrt{3.5}\right)$ and "cat" $\left(|\alpha_L| \geq \sqrt{3.5}\right)$. (see Supplementary Table 1 and Supplementary Note 3)

To demonstrate the coherence property of the a-CS, the interference between the input a-CS (in SLM_G, $|\alpha_L\rangle + |-\alpha_L\rangle$) and a rotated state (in SLM_M, $|\alpha_L\rangle + e^{i\theta}|-\alpha_L\rangle$) was measured in the equatorial plane of the Bloch sphere (see Fig. 4a). This type of interference directly evaluates the coherence property of the generated a-CS. Figure 4b presents two interference curves of the odd (red) and even (blue) a-CS of $|\alpha_L| = \sqrt{3.5}$, where the $x$-axis is the relative phase angle of the rotation state in SLM_M and the $y$-axis is the corresponding power. Additionally, one can evaluate the coherence by calculating the visibility ($V = P_{\mathrm{Max}} - P_{\mathrm{Min}}/P_{\mathrm{Max}} + P_{\mathrm{Min}}$). In Fig. 4c, four orange bars indicate the average visibilities of four a-CSs from "kitten" to "cat". The clear interference pattern and high visibility show that the coherence of an a-CS is satisfactory and usable.

To study the wavefront of the a-CS, we probed the interference between the sum of the superposed "wave packets" of the a-CS and a referenced plane wave. The theoretical and experimental interference patterns for the a-CS $\left(|\alpha_L| = \sqrt{3.5}\right)$ are presented in Fig. 4d, e, respectively. The detailed calculations are presented in the methods section, and other results can be found in Supplementary Note 4. An off-axis interferogram was created by an input field and a reference plane wave with a small incident angle[42]. The type of interference seen in the resulting patterns demonstrates that the two separated GILs have two genuine Gaussian wavefronts without any screw dislocation.

Finally, we evaluated the a-CS using standard state tomography with mutually unbiased bases (MUBs), which is an exhaustive and reliable tool to estimate a high-dimensional quantum state[43]. These MUBs can be generated by a discrete Fourier transformation in a dimension of prime order[44] (also see Supplementary Note 5). Several groups have achieved the reconstruction of high-dimensional superposed classical states from $d=2$ to 6 in OAM space[45,46], where the similarities are calculated between experimental and theoretical projections in all MUBs. Using this reliable strategy, we conducted the state tomography on odd "kitten" $\left(|\alpha_L|=\sqrt{0.5}\right)$ and "cat" $\left(|\alpha_L|=\sqrt{5.0}\right)$ states in OAM space with dimensions of 7 and 17, respectively. Figure 5a, b shows the theoretical projection- matrix calculated in all MUBs, and Fig. 5c, d show the results of the experimental measurements. The projection matrices yielded similarities of $S=0.9907$ and 0.9533 for the analogous "kitten" and "cat" states, respectively. In principle, one could also calculate the density matrix and fidelities[47]; however, we will not discuss them further in this paper as the similarity is an analogous quantity to fidelity[45]. The high similarities and the visualization of the spatial projection-matrix directly demonstrate that the a-CSs generated in our system are reliable and are very similar to the ideal cat-like state.

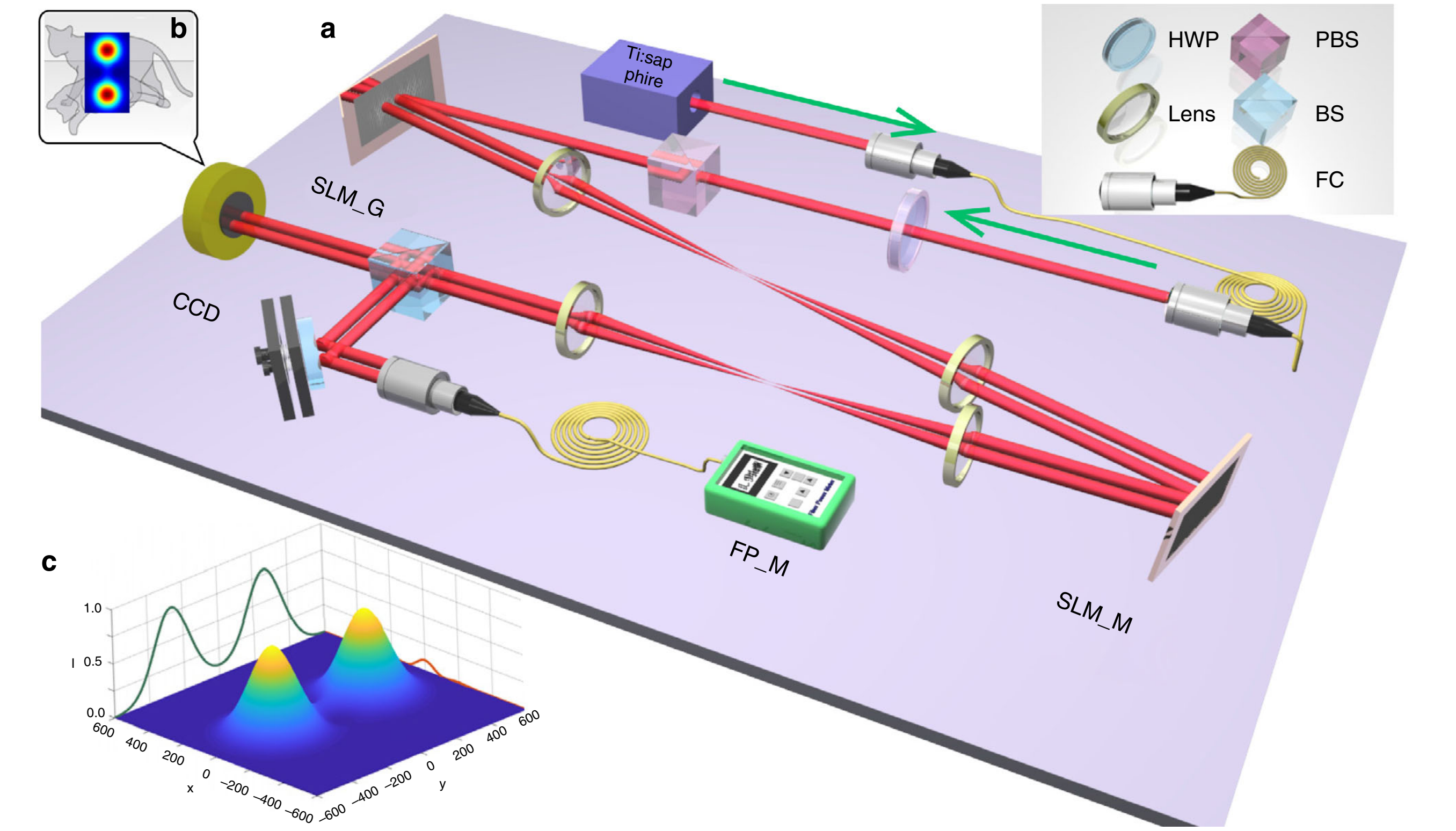

**Fig. 2** Schematic of the experimental setup for engineering the analogous cat state (a-CS). **a** Detailed illustration of the generation and measurement of an a-CS. **b** Illustration of the intensity -distributions of an a-CS with corresponding "dead" and "alive" cat[41]. **c** The three-dimensional intensity -distributions for an even cat state. SLM: spatial light modulator. HWP: half wave plate, (P)BS: (polarization) beam splitter, FC: fiber coupler. During the simulation process, the maximum OAM mode was 50

**The effect of atmospheric turbulence on a-CS**. A q-CS is extremely fragile owing to due to "decoherence". Similarly, "decoherence" can also occur for an analogous OAM-based cat state during its manipulation[48]. In this section, we quantify the effects of atmospheric turbulence on the coherence of OAM-based cat states. Research on the effect of turbulence of OAM light is a valuable topic in both the classical and quantum communications fields. Various methods have been proposed to simulate the effects of atmospheric turbulence on OAM light, for instance, the multiple random phase screen (MRPS)[49–51], and rotation coherent function methods[52]. Here, we simulated the "decoherence" of several OAM eigenstates and the a-CS using the MRPS method[49].

When a coherent monochromatic OAM light beam $|\psi\rangle_{in}$ is transmitted through a turbulent medium, one can estimate the output state $|\psi\rangle_{out}$ and measure the OAM spiral spectrum. In the MRPS method, there are three steps to simulate the output fields. First, one should give an initial field $|\psi\rangle_{\rm in}$, which may be a single or a superposition OAM state. Second, the total propagation distance $Z$ is divided into $N$ segments, and the field $|\psi\rangle_{N=j}$ after turbulence is calculated from one screen $V_{j-1}(r)$ to the next screen $V_j(r)$. Finally, one obtains the final field $|\psi\rangle_{N=N}$ and measures the spiral spectrum (probability) $P(L|\psi)=\langle L|\psi\rangle_{\rm out}=\left|\int\varphi_L(r,\theta)^{+}\psi_{\rm out}(r,\theta)r{\rm d}r{\rm d}\theta\right|^2$. The transmission progress between two phase screens can be described by a stochastic parabolic equation[53,54];

$$-2ik\frac{\partial}{\partial z}\psi(r)=\nabla_T^2\psi(r)+2k^2V(W,r)\psi(r) \quad (4)$$

where $k=2\pi/\lambda$ is the wavenumber; $\nabla_T^2=\partial_x^2+\partial_y^2$ represents a two-dimensional Laplacian in the transverse plane; and $V(W,r)$ gives the phase and intensity fluctuation controlled by the power spectral density[50,51,54], where $W$ is the turbulence strength depended on the Fried parameter $r_0$ that associates with propagation distance and refractive index structure constant[55]. We solved Eq. (4) using the split-step Fourier method[53,54].

Based on the MRPS method, we simulated the propagation process of an a-CS passing 1 km through Kolmogorov turbulence ($W=2.0$)[49,54,56,57]. Figure 6a, b shows the simulation results of the intensity- and phase- distributions for an a-CS, and the corresponding theoretical distribution can be found in Fig. 1. We found that the profiles of the two GILs and their phase distribution remained intact. In fact, the output beam was still a transverse coherence field, and it could be decomposed into LG

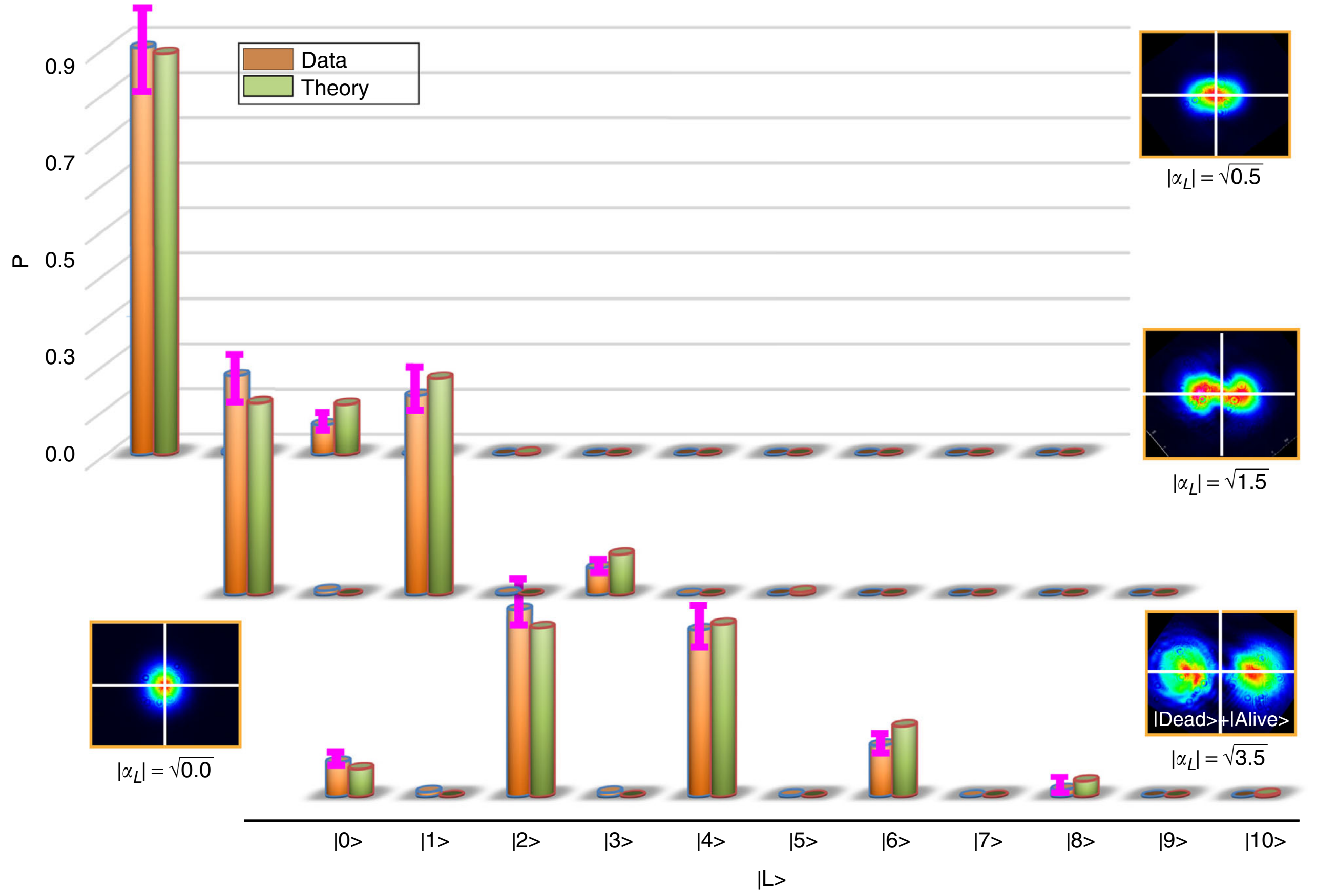

**Fig. 3** Spatial mode- and intensity- distributions of the analogous cat state (a-CS) for the "kitten" and "cat" state. In each plot, the orange and green bars represent the spatial mode- distributions for $|\alpha_L| = \sqrt{0.5}, \sqrt{1.5}, \text{and}\sqrt{3.5}$ that were experimentally measured and calculated, respectively, where $|\alpha_L|$ represents the size of a-CS. The *x*-axis represents the orbital angular momentum eigenstates loaded in spatial light modulator (SLM_M), while the *y*-axis is the normalized power measured by the fiber power meter. The pink error bars represent the experimental error bars during the measurement. The images acquired by a charge coupled device (CCD) record the corresponding intensity of the cat state, and the image inserted in the left bottom shows a Gaussian beam as a reference

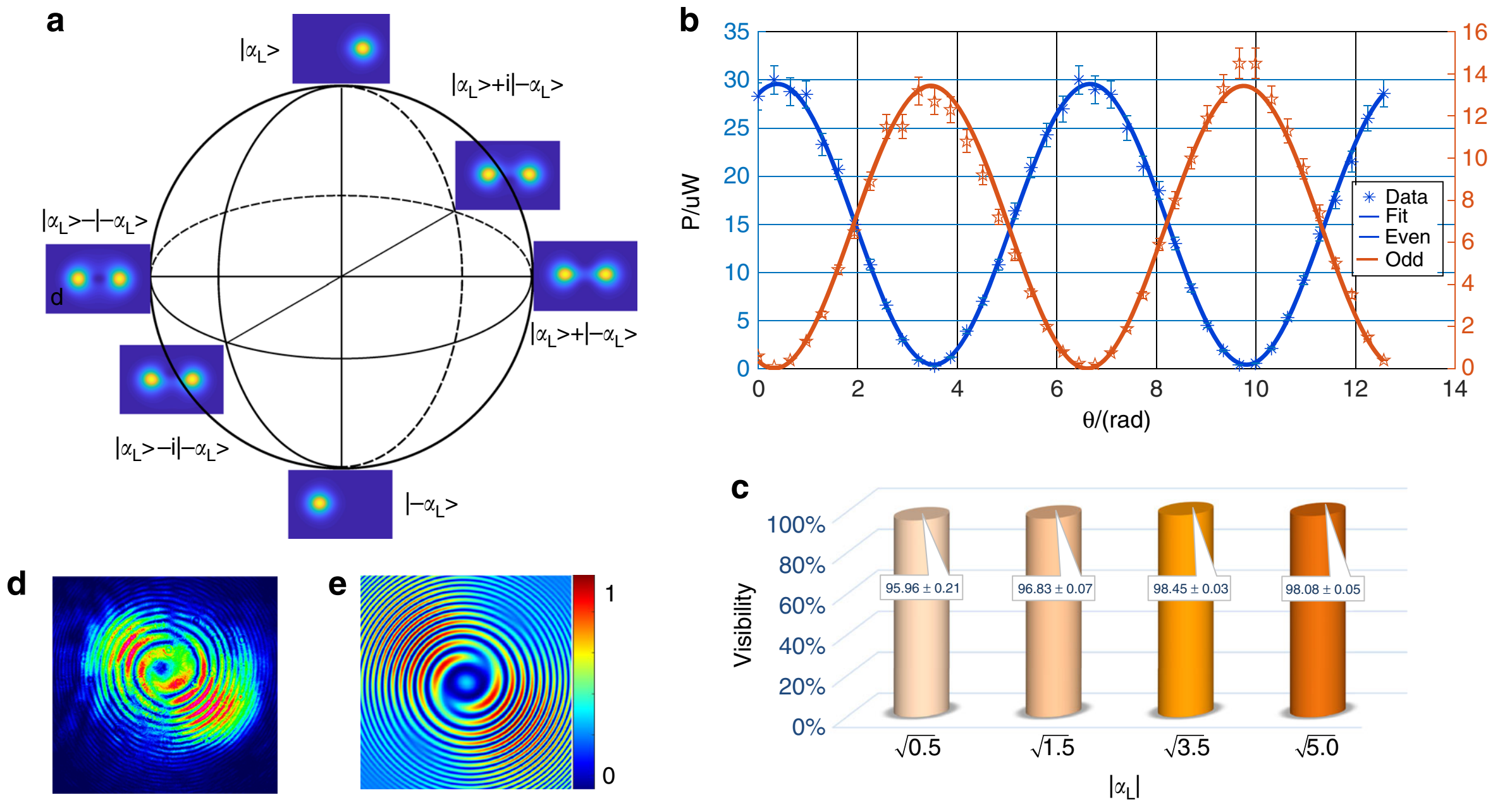

**Fig. 4** Coherence of the analogous cat state (a-CS). **a** The Bloch-sphere representation of an a-CS with a size of $|\alpha_L| = \sqrt{1.5}$. **b** Interference between even (blue line) and odd (red line) a-CS $(|\alpha_L\rangle \pm |-\alpha_L\rangle)$. **c** The visibilities for analogous "kitten" and "cat" states. **d** and **e** The measured and theoretical interference patterns between an a-CS $(|\alpha_L| = \sqrt{3.5})$ and a referenced plane wave (magnified Gaussian beam), where $|\alpha_L|$ represents the size of a-CS

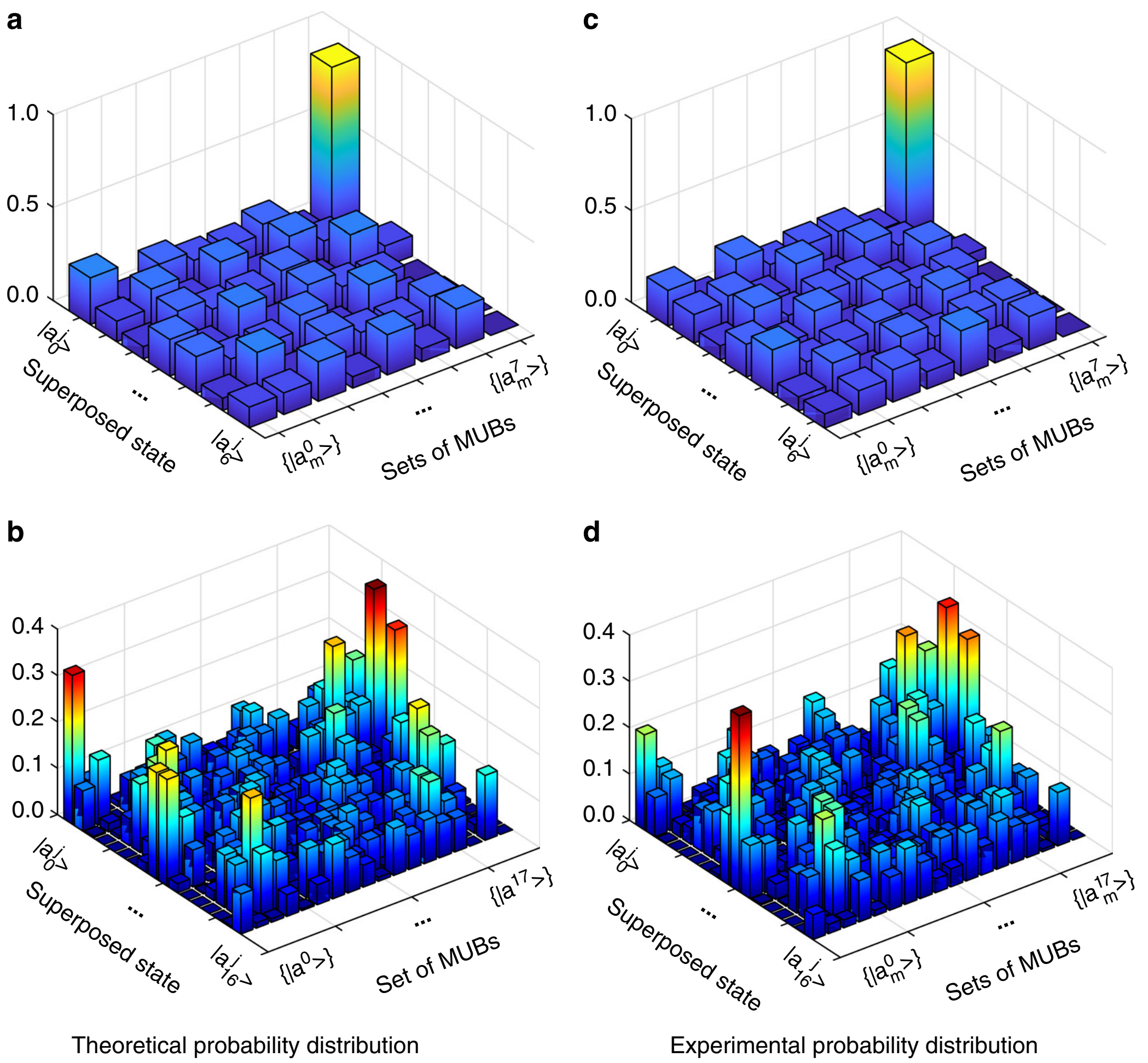

**Fig. 5** Experimental projections of the analogous "kitten" $\left(|\alpha_L| = \sqrt{0.5}\right)$ and "cat" $\left(|\alpha_L| = \sqrt{5.0}\right)$ state in dimensions 7 and 17. **a** and **b** Three-dimensional representation of the normalized theoretical spatial distribution matrices for the "kitten" $\left(|\alpha_L| = \sqrt{0.5}\right)$ and "cat" $\left(|\alpha_L| = \sqrt{5.0}\right)$ state, where $|\alpha_L|$ is the size of analogous cat state. **c** and **d** show the corresponding normalized experimental distributions. In these panels, the *x*-axis corresponds to sets of mutually unbiased bases (MUBs) from 0 to dimension d. Meanwhile, the *y*-axis corresponds to the superposition states for each set of MUBs and the *z*-axis gives the normalized power distribution measured by power meters for each basis

modes (mode dispersion). For example, a single OAM state $|L_0\rangle$ can be written as $\sum_{\Delta L=-\infty}^{+\infty} C_{L_0}^{\Delta L}|L_0 + \Delta L\rangle$ after the turbulence. For an a-CS, i.e., an even cat state that only occupies even modes, the modes will couple to odd modes under turbulence, which leads to distortions between them. To describe the distortions, we employed the fidelity $F = Tr\left[\sqrt{\sqrt{\rho}\rho_{\mathrm{sim}}\sqrt{\rho}}\right]^2$ between the in- and outputs, where $\rho_{\mathrm{sim}}$ refers to the simulated density matrix in a $d$-dimensional subspace of the total OAM Hilbert space, and $\rho$ belongs to the theoretical density matrix. The corresponding results are shown in Fig. 6c, where blue and red lines represent the fidelities of the "kitten" and "cat" states. The fidelity of the "cat" state falls faster than that of the "kitten" state, which illustrates that the large cat state is more fragile than the smaller one. Figure 6d shows the normalized detected probabilities of the OAM eigenstate with $L = 0$, 10, and 20, respectively. In Fig. 6c, d, each data point represents an average value of the results of ten simulations for the same turbulence strength. We found that the higher the OAM mode light was, the faster "decoherence" appeared. The simulation results given in our calculation are the same as those in ref. [52] using the rotation coherent function. Because of the mode dispersion, the a-CS appears as if "decoherence" had occurred. However, the mode distortion induced by turbulence can be recovered by using adaptive optics[53,54]. For visualization, we make a flash in Supplementary Movie 3 to present the overall dynamic process, including random phase, distributions for intensity and phase, and the spiral spectrum.

## Discussion

To summarize, we have taken the first step toward simulating a well-controlled a-CS using spatial OAM modes. In photonic OAM space, the a-CS is easy to create, manipulate, and measure, and has a good coherence. In the preceding sections, the scheme proposed for the generation of the cat states has many advantages, the main two being that the size of the cat states can be expanded in the laboratory using linear optical elements and that the a-CS has the ability to simulate the behaviors of optical q-CS in phase space, i.e., movement, rotation, and interference. Therefore, the a-CS provides an excellent "laboratory" in the classical world to study the behaviors of optical cat states in the quantum world.

In the future, the a-CS can be easily expanded to the single-photon level. On the one hand, the cat state can be post-created via changing the inputs, for instance, using a heralded

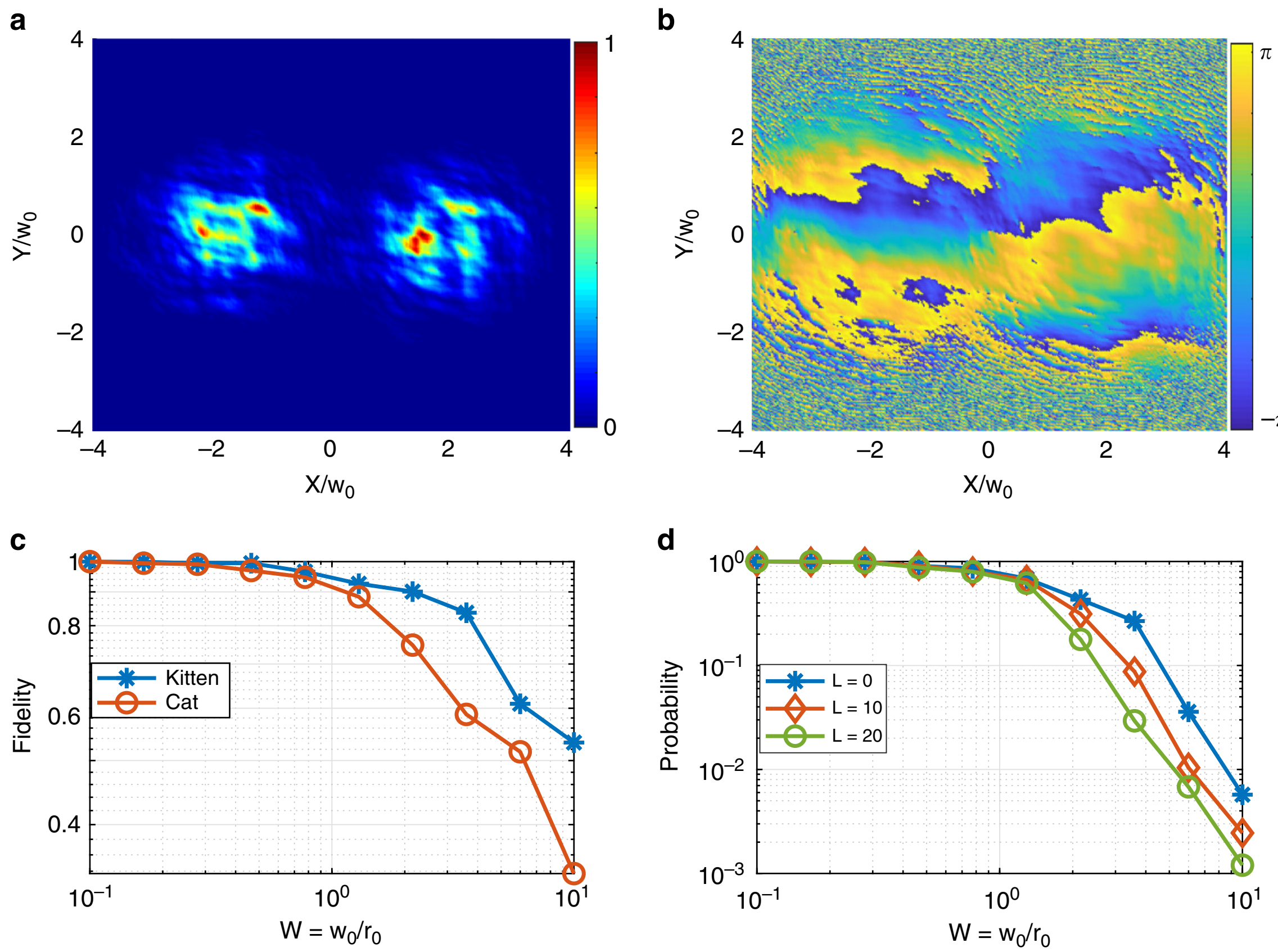

**Fig. 6** Simulation results for orbital angular momentum (OAM) states passing through atmospheric turbulence. **a**, **b** Intensity- and phase- distributions for an analogous cat state $\left(|\alpha_L|=\sqrt{3.5}\right)$, where $x(y)$-axis corresponds to the position coordinate against the beam waist ($w_0$), and $|\alpha_L|$ represents the size of the cat state. **c** Fidelities of "kitten" and "cat" states between the in- and outputs under Kolmogorov turbulence. **d** Normalized detected probability for the pump in single OAM states. In panels **c** and **d**, the $x$-axis represents the turbulence strength depended on the Fried parameter $r_0$ and beam waist. In all simulations, the waist of the Gaussian beam was 0.0735 m; each segment was located 2 m between two screens, and the number of Fourier mode was 256

single-photon source. On the other hand, it can be directly generated by the process of spontaneous parametric down conversion[58], with the help of the adaptive shaping technology[59,60]. Because the cat state is encoded in spatial mode space with good coherence, the proposed source will support a new platform for studying cat state-based protocols, for example, the generation of entangled coherent states (Bell-cat states) $N_{\pm}(|\alpha\rangle|-\alpha\rangle\pm|-\alpha\rangle|\alpha\rangle)$ in OAM space with the help of a two-dimensional entangled source. The entangled coherent state is a vital resource for cat state-based protocols, i.e., quantum teleportation and computation[4,61]. Recently, some fundamental high-dimensional quantum gates in OAM space have been constructed[62–64]. Furthermore, the original scenario presented by Schrödinger's cat describes an entangled system between a cat and an atom, i.e., $|\alpha\rangle|\uparrow\rangle+|-\alpha\rangle|\downarrow\rangle$, where $|\alpha\rangle$ signifies the state of an alive or dead cat, and $|\uparrow,\downarrow\rangle$ signifies the internal state of the atom[65,66]. The thought experiment can be demonstrated easily in our regime; specifically, $|\uparrow,\downarrow\rangle$ can be employed by the spin angular momentum (polarization) DOF of the photon. Additionally, the ability to generate such states is promising as it may allow us to explore the physics associated with cat states and simulate some new quantum Gedankenexperiments, such as those proposed in the recent debate about the extended Wigner's Friend Paradox (Hardy's paradox)[67].

## Methods

**Experimental details to achieve the cat states**. For generating the cat states studied in this paper, the main technological challenge lies in how to effectively manipulate many asymmetric OAM modes, i.e., $\sum_{L=0}^{50} c_L|L\rangle$. Most previous studies were only able to demonstrate an asymmetric superposition of several OAM modes[46,68]. In our regime, we successfully generated a superposed state of an arbitrary number of modes by revising the beam waists. The SLM was the only phase-modulation element used in the experiment, and each pixel could be imprinted with a gray map to modulate the phase of the incoming light (between 0 and $2\pi$). We loaded the amplitude-encoded phase hologram onto the SLM to generate an a-CS[45,46,69]. For a single OAM eigenstate $|L\rangle$ associated with an LG mode, $LG_p^l(\rho,\phi)$, the phase profile imprinted in the SLM is given by:

$$\Phi(x,y)_{\text{holo}} = -\text{sinc}(\pi M-\pi)\cdot \text{Arg}(E_l(x,y))\cdot \Lambda(x,y)$$

where $M$ $(0\le M\le 1)$ is the normalized bound for the positive function of the amplitude; $E_l(x,y)$ is an analytical function of the amplitude and phase profiles of the desired OAM field, and $\Lambda(x,y)$ is the optical blazing phase for generating pure OAM states[69]. To get the phase hologram of the a- CS, we replaced the electric field $E_l(x,y)$ with the desired superposed field in Eq. (1).

For the LG mode, the beam waist $w(z)$ relates to the topological charge $L$. The standard deviation of the spatial distribution of the beam $r_{\text{rms}}$ illustrates the divergence of the OAM fields[70]. In the plane of the beam waist of $z=0$, this radius should be $r_{\text{rms}}(0)=\sqrt{|L|+1/2}\cdot w_0$, where $w_0$ is the beam waist of the input Gaussian beam. Thus, the hologram profiles of a large-$L$ mode may exceed the size of the SLM, so we need to adjust the beam waist radius to match the input Gaussian beam. To match the diameters of the different illuminating beams, Kotlyar et al. state that the waist of the Gaussian beam and the radius of the circular aperture should be matched by making the first momenta of their intensity distribution, which gives rise to the analytic expression in ref. [71]. We implemented the inverse of this process and modified the beam waist $w_L(z=0)$ to

$$w_L(z=0)=\frac{2}{\sqrt{\pi}}w_0\left[\frac{\sqrt{\pi}}{|L|!}2^{-(|L|!+1)}\prod_{n=0}^{|L|}(2n+1)\right]=w_0\frac{2^{-|L|}(2|L|+1)!!}{|L|!} \quad (5)$$

where $n!!=1\times 3\times 5\ldots\times n$ and $L$ is the topological charge. In this situation, although the beam waists differ for a series of OAM modes, the orthogonality

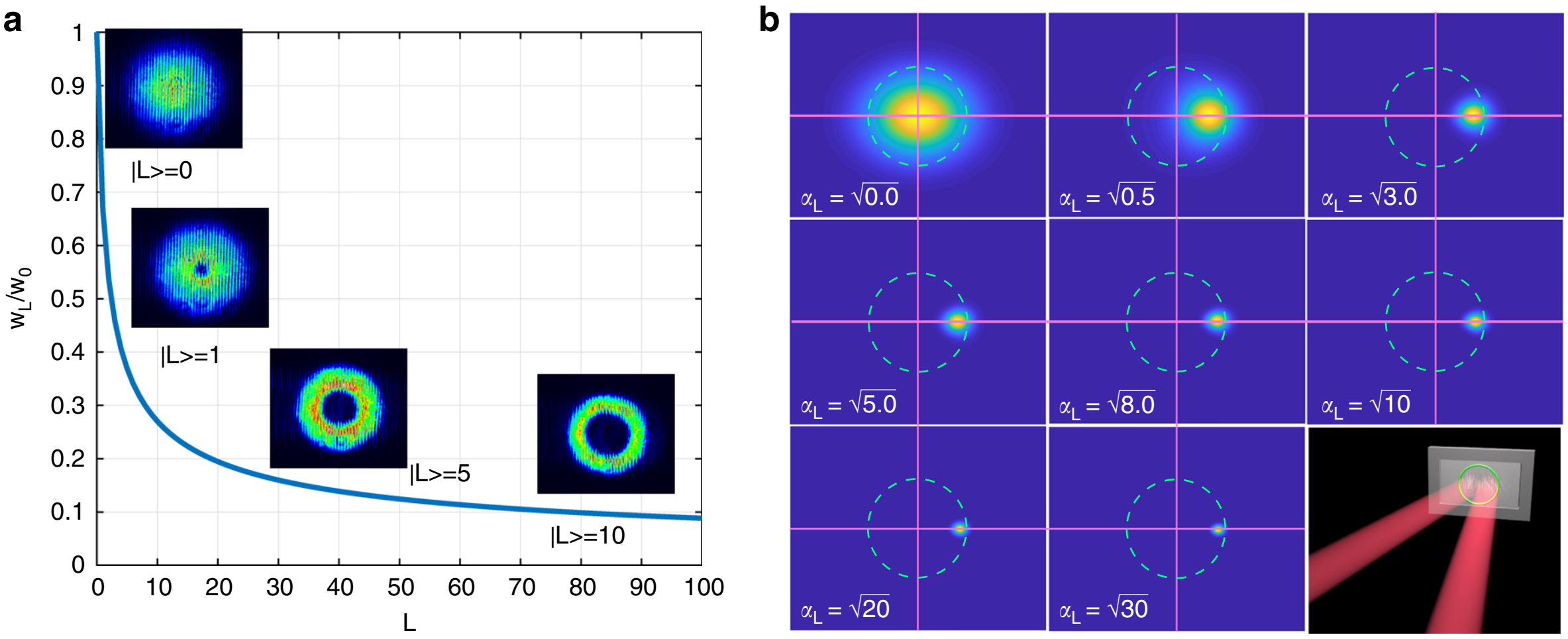

**Fig. 7** Intensity distribution for the orbital angular momentum (OAM) states using the revised Laguerre-Gaussian mode. **a** Beam waist ratio vs. topological charge $L$ based on Eq. (5); the four inset graphics show the corresponding intensity of states $L$ (=0, 1, 5, 10) recorded by the charge coupled device (CCD) camera. **b** Normalized intensity distribution of analogous coherent states from 0 to, where the wavelength, beam waist, and maximum OAM mode were 0.780 μm, 200 μm, and 100, respectively, in the simulations

between them still holds true[72]. Figure 7a shows the relationship between the waist ratios ($w_L/w_0$) and the OAM eigenstates; insets are several intensity distributions measured in the focus plane of the 4$f$ system. Figure 7b shows the situation of the a-CST as it evolves from 0 to $\sqrt{30}$, where the beam waists $w_L$ for different OAM modes obey Eq. (5). From Fig. 7b, one can see that the beam size of the a-CST decreases as the size of the a-CST increases and that the movement of the beam gradually reaches its limit (indicated by the green circle), which is set by the input Gaussian beam. Therefore, it is inevitable that a large a-CS has a low reflection efficiency in its generation.

**Details for testing the wavefront of the a-CS**. To study the wavefront of the a-CS, we studied the interference between the sum of the GILs and a referenced beam. For the single pure vortex fields, the interference pattern was a Fork type grating[42,73], which depicts the wavefront with a screw dislocation. The setup for testing the wavefront of the a-CS can be found in Supplementary Fig. 5, where a CCD was used to observe the interference patterns in port C or D. The theoretical interference patterns were calculated according to:

$$I_{\mathrm{Inter}} = |\langle \mathrm{Cylin}(r,\phi,z)|Cat\rangle + a\cdot\langle \mathrm{Cylin}(r,\phi,z)|P\rangle|^2 \quad (6)$$

where $\langle \mathrm{Cylin}(r,\phi,z)|\mathrm{P}\rangle (= E_0\cdot\exp(-i(kz+y\sin(\theta))))$ can be regarded as a plane wave and $a$ is the attenuation factor of the neural attenuator. $\langle \mathrm{Cylin}(r,\phi,z)|\mathrm{Cat}\rangle$ is the electric field at the center of the BS, which can be calculated using the Collins diffraction integral formula[74]. Other simulations and further results for the cat states can be found in the Supplementary Fig. 6.

## Data availability

The data that support the findings of this study are available from the corresponding author upon reasonable request.

## Acknowledgements

We thank Prof. Alexander Lvovsky and Barry C. Sanders of the University of Calgary for discussing the fundamental advantages and coherence of cat states. We also thank Prof. Vlatko Vedral of the University of Oxford for fruitful discussions; Prof. Uwe R. Fischer of the Seoul National University for discussing the stability of the cat state, and Dr. Kun Huang of the University of Shanghai Science and Technology for giving some suggestions to improve the quality of paper. This work is supported by The Anhui Initiative in Quantum Information Technologies (AHY020200); National Natural Science Foundation of China (61435011, 61525504, 61605194, 61775025, 61405030); China Postdoctoral Science Foundation (2016M590570,2017M622003); Fundamental Research Funds for the Central Universities; National Key R&D Program of China (2018YFA0307400).

## Author contributions

S.-L.L. conceived and designed the scheme for the classical analogy of a cat state; S.-L.L., Q.Z., Z.-Y.Z., and S.K.L. performed the experiment with assistance from Y.-L. and Y.-H. L. S.-L.L. wrote the draft manuscript with the help from B.-S. S. Q.Z. polished the language in English. All authors discussed the results, and implications, and commented on the manuscript. Z.-Y.Z., G.-C.G, and B.-S.S. supervised the project.

## Additional information

**Supplementary information** accompanies this paper at https://doi.org/10.1038/s42005-019-0156-2.

**Competing interests:** The authors declare no competing interests.

**Reprints and permission** information is available online at http://npg.nature.com/reprintsandpermissions/